\documentclass[preprint,showpacs,fleqn,nobibnotes]{revtex4}

\usepackage{amsmath}
\usepackage{graphicx}
\usepackage{float}
\usepackage{subfigure}

\newcommand{\be}{\begin{equation}}
\newcommand{\ee}{\end{equation}}

\def\lsim{\raise0.3ex\hbox{$<$\kern-0.75em\raise-1.1ex\hbox{$\sim$}}}
\def\gsim{\raise0.3ex\hbox{$>$\kern-0.75em\raise-1.1ex\hbox{$\sim$}}}

\def\pom{{I\!\!P}}

\def\beq{\begin{equation}}
\def\eeq{\end{equation}}
\def\bea{\begin{eqnarray}}
\def\eea{\end{eqnarray}}
\def\bq{\begin{quote}}
\def\eq{\end{quote}}

\newcommand{\rr}{\mbox{$r$}}

\newcommand{\rkr}{\mbox{\boldmath $r$}}

\def\gappeq{\mathrel{\rlap {\raise.5ex\hbox{$>$}}
{\lower.5ex\hbox{$\sim$}}}}

\def\lappeq{\mathrel{\rlap{\raise.5ex\hbox{$<$}}
{\lower.5ex\hbox{$\sim$}}}}

\def\Toprel#1\over#2{\mathrel{\mathop{#2}\limits^{#1}}}

\def\lsim{\raise0.3ex\hbox{$<$\kern-0.75em\raise-1.1ex\hbox{$\sim$}}}
\def\gsim{\raise0.3ex\hbox{$>$\kern-0.75em\raise-1.1ex\hbox{$\sim$}}}

\def\pom{{I\!\!P}}
\newcommand{\rk}{\mbox{\boldmath $k$}}

\begin{document}

\title{Probing QCD dynamics in  two-photon interactions at high energies}
\author{ V.P. Gon\c{c}alves $^{1}$, M.V.T. Machado  $^{2}$ and W. K. Sauter $^{3}$ }

\affiliation{$^{1}$ Instituto de F\'{\i}sica e Matem\'atica,  Universidade
Federal de Pelotas\\
Caixa Postal 354, CEP 96010-900, Pelotas, RS, Brazil\\
$^{2}$ \rm Centro de Ci\^encias Exatas e Tecnol\'ogicas, Universidade Federal do Pampa \\
Campus de Bag\'e, Rua Carlos Barbosa. CEP 96400-970. Bag\'e, RS, Brazil \\
$^{3}$ \rm High Energy Physics Phenomenology Group, GFPAE  IF-UFRGS \\
Caixa Postal 15051, CEP 91501-970, Porto Alegre, RS, Brazil}

\begin{abstract}
In this paper the  two-photon interactions at high energies  are investigated considering different approaches for the QCD dynamics. In particular, we calculate the $\gamma^* \gamma^*$ total cross section in different theoretical approches and present a comparison among the predictions of the BFKL dynamics at leading and next-to-leading order with those from saturation physics. We analyze the possibility that the future linear colliders could discriminate between these different approaches.
\end{abstract}

\maketitle

The high energy limit of the perturbative QCD is characterized by a center-of-mass which is much larger than the  hard scales present in the problem. The simplest process
where   this limit can be studied  is the high energy scattering
between two heavy quark-antiquark states, {\it i.e.} the
onium-onium  scattering. For a sufficiently heavy onium state,
high energy scattering is  a perturbative process since the onium
radius gives the essential scale at which the running coupling
$\alpha_s$ is evaluated. In the dipole picture \cite{dipole},
the heavy quark-antiquark pair and the soft gluons in the limit of
large number of colors $N_c$ are viewed as a collection of color
dipoles. In this case, the cross section can be understood as a
product of the number of dipoles in one onium state, the
number of dipoles in the other onium state and  the  basic cross
section for dipole-dipole scattering due to two-gluon exchange. At
leading order (LO),  the cross section grows rapidly with the
energy ($\sigma \propto \alpha_s^2  \, e^{(\alpha_{\pom} - 1)Y}$, where
$(\alpha_{\pom} - 1) = \frac{4\alpha_s\,N_c}{\pi}\,\ln 2 \approx 0.5$ and $Y =
\ln\,s/Q^2$) because the LO BFKL equation \cite{bfkl} predicts that the number of dipoles in the light cone
wave function grows rapidly with the energy. Several shortcomings are present in this calculation. Firstly, in the  leading order calculation  the energy scale is arbitrary, which implies that the absolute value to the total cross section is therefore not predictable. Secondly, $\alpha_s$ is not running at LO BFKL. Finally, the power growth with energy violates $s$-channel unitarity at large rapidities. Consequently, new physical effects should modify the LO BFKL equation at very large $s$, making the resulting amplitude unitary.

 A theoretical possibility to modify this behavior in a way consistent with the unitarity is the idea of parton saturation \cite{BK}, where non-linear effects associated to high parton density are taken into account. The basic idea is that when the parton density increases (and the scattering amplitude tends to   the unitarity limit), the  linear description present in the BFKL equation breaks down and one enters the saturation regime, where the dynamics is described by a nonlinear evolution equation and the parton densities saturate \cite{BK}.  The transition line between the linear and nonlinear regimes is characterized by the saturation scale $Q_{\mathrm{sat}} (x)$, which is energy dependent and  sets the
critical transverse size for the unitarization of the cross sections. The saturation approach implies that the onium-onium cross section still grows with the energy, but the rise is slower than $s^{0.5}$ \cite{salam_onium}.
In Ref. \cite{mueller}, Mueller has argued that these corrections become important at the values of rapidity of the order of $Y_U \approx \frac{1}{\alpha_{\pom} - 1} \ln \frac{1}{\alpha_s^2}$.

Another possible solution, which is expected to diminishes the energy growth of the total cross section, is the calculation of higher order corrections to the BFKL equation. After an effort of ten years, the next-to-leading order (NLO) corrections  were  obtained \cite{bfklnlo1}  and the spurious singularities were solved considering renormalization-group improved regularizations (for a review on
NLO BFKL corrections, see e.g. Ref. \cite{nlosalam}  and
references therein). In particular, in Ref. \cite{NLOBLM}  the Brodsky-Lepage-Mackenzie (BLM) optimal scale setting procedure \cite{BLM} was used to eliminate the renormalization scale ambiguity present in the $\overline{\mbox{MS}}$-scheme calculations. The NLO BFKL predictions, as improved by the BLM scale setting, yields $(\alpha_{\pom} - 1) = 0.13 - 0.18$. In Ref. \cite{kovmue} the authors have shown that due to the running coupling effects the NLO corrections become important at the rapidities of the order of $Y_{NLO} \approx \alpha_s^{-5/3}$. Consequently, in principle, one has $Y_U \ll Y_{NLO}$ at parametrically small $\alpha_s$. That implies that the center of mass energy at which the saturation effects become important  is much smaller than the energy at which the NLO corrections start playing an important role. However, it still is an open question, since it only can be definitively established once next-to-leading-logarithmic contributions are fully computed for the non-linear evolution equations present in the saturation approaches. One possibility to check this assumption is  the analysis of the energy dependence of the onium-onium cross section, which  could disentangle the QCD dynamics.

A reaction which is analogous to  the process of scattering of two onia discussed above is the off-shell photon scattering at high energy in $e^+\,e^-$ colliders, where
the photons are produced from the leptons beams by bremsstrahlung (For a review see, e.g., Ref. \cite{nisius}). In these two-photon reactions, the photon virtualities can be made
large enough to ensure the applicability of the perturbative
methods or can be varied in order to test the transition between the soft and hard regimes of the QCD dynamics. 
From the view of the BFKL approach, there are several calculations using the leading
logarithmic  approximation \cite{bartels,gamboone}  and
 considering some of the next-to-leading corrections to the total
cross section $\gamma^* \gamma^*$ process
\cite{gamboone,gammaNLO}. In particular, the  stable next-to-leading order  program
relying on the  BLM optimal scale setting \cite{NLOBLM} has
produced  good results with OPAL and L3 data at LEP2
\cite{gammaNLO}. 
On the other hand,  the successful description of all
inclusive and diffractive deep inelastic data at the collider HERA
by saturation models \cite{GBW,bgbk}  suggests that these effects might become important
in the energy regime probed by current colliders. It motivated the generalization of  the saturation model  to two-photon interactions at high energies performed in Ref. \cite{Kwien_Motyka},   which has obtained a very good description of the data on the $\gamma \gamma$ total cross section, on the photon structure function at low $x$ and on the $\gamma^* \gamma^*$ cross section.
The formalism used in Ref. \cite{Kwien_Motyka} is based on the dipole picture \cite{dipole}, with the  $\gamma^* \gamma^*$ total cross sections 
being described  by the interaction of two color dipoles, in which the virtual photons fluctuate into (For previous analysis using the dipole picture see, e.g.,  Refs. \cite{nik_photon,dona_dosch}). The dipole-dipole cross section is modeled considering phenomenological models based on saturation physics. A current shortcoming of the saturation approaches is that the non-linear evolution equations (BK and JIMWLK) were derived at leading order accuracy, resumming all powers of $\alpha_s \ln s$ along with all the multiple scatterings in the target, but disregarding for instance any running coupling corrections in their kernels (For recent efforts see, e.g. \cite{kovwei_alfa}). Another important aspect which should be emphasized is that these evolution equations contains the LO BFKL equation in the low density limit.

As the NLO BFKL predictions as well as the saturation one describes quite well the LEP data \cite{opal}, a current open question is if the 
future $e^+ e^-$ colliders  will allow to discriminate
between BFKL and saturation predictions. The main goal of this letter is to present a  comparison of the (LO and NLO) BFKL  and saturation predictions  for the energy dependence of the total cross section and verify if it is possible discriminate between these approaches in the kinematical range of the future linear colliders.

Lets start presenting the main formulas to calculate the total cross section in the BFKL formalism. one has that the  total cross section of two unpolarized gammas with virtualities $Q_{1}$
and $Q_2$ in the LO BFKL formalism reads as follows:
\begin{eqnarray}
\sigma(s_{\gamma \gamma},Q_1^2,Q_2^2) =
  \sum_{i,k = T,L}
\frac{1}{\pi \sqrt{Q_1^2 Q_2^2}} \int_{0}^{\infty}
\frac{d \nu}{2 \pi} 
 \cos \Biggl[\nu \ln \biggl(\frac{Q_1^2}{Q_2^2}\biggr)\Biggr] 
\Phi_{i}(\nu) \Phi_{k}(-\nu)
\Biggl( \frac{s}{s_0}\Biggr)^{\omega(Q^2,\nu)},  
\label{eqn:sigma-g}
\end{eqnarray}
where one has used as kinematic variables the $\gamma^* \gamma^*$ c.m.s. energy
squared $s_{\gamma \gamma}=W_{\gamma \gamma}^2=(p+q)^2$, with $p$ and $q$ being  the
photon momenta and  the photon virtualities squared  given by 
$Q_1^2=-q^2$ and $Q_2^2 = -p^2$. Moreover, 
$\Phi_{i}(\nu)$ are the transverse and longitudinal polarizations gamma impact factors, which at LO accuracy
 are given by :
\begin{eqnarray}
\Phi_T(\nu) = \Phi_T(- \nu) = \alpha \, \alpha_S \, 
\Bigg( \sum_q e_q^2 \Bigg)  \frac{\pi}{2}
\frac{\Bigl[\frac{3}{2} - i \nu \Bigr] 
\Bigl[\frac{3}{2} + i \nu \Bigr]
\Gamma \Big(\frac{1}{2} - i \nu \Big)^2
\Gamma \Big(\frac{1}{2} + i \nu \Big)^2}
{\Gamma (2 - i \nu) \Gamma (2 + i \nu)} 
\label{eqn:impactT}
\end{eqnarray}
and
\begin{eqnarray}
\Phi_L(\nu) = \Phi_L(- \nu) = \alpha \, \alpha_S \,
\Bigg( \sum_q e_q^2 \Bigg) \pi
\frac{\Gamma \Big(\frac{3}{2} - i \nu \Big) 
\Gamma \Big(\frac{3}{2} + i \nu \Big) 
\Gamma \Big(\frac{1}{2} - i \nu \Big)
\Gamma \Big(\frac{1}{2} + i \nu \Big)}
{\Gamma (2 - i \nu) \Gamma (2 + i \nu)}. 
\label{eqn:impactL}
\end{eqnarray}
The Regge scale parameter $s_0$ is proportional to
a hard scale $Q^2 \sim Q_1^2,Q_2^2$; $\Gamma$ is the Euler 
$\Gamma$-function and $e_q$ is the quark electric charge.

The high-energy asymptotics of the cross sections is determined by 
the highest eigenvalue, $\omega^{max}$, of the BFKL 
equation \cite{bfkl}: $\sigma \sim 
s^{\alpha_{\pom}-1} = s^{\omega^{max}}$. At leading order it is  rather large: 
$\alpha_{\pom} - 1 =\omega_L^{max} = 
12 \, \ln2 \, ( \alpha_S/\pi )  \simeq 0.55 $ for 
$\alpha_S=0.2$. On the other hand, the eigenvalue of the  NLO BFKL equation at transferred momentum 
squared $t=0$ in the
 $\overline{\mbox{MS}}$-scheme reads
\begin{equation}
\omega _{\overline{MS}}(Q_{1}^{2},\nu ) =
 N_c \chi_{L}(\nu ) \frac{\alpha_{\overline{MS}}(Q_{1}^{2})}{\pi }
\Biggl[ 1 + r_{\overline{MS}}(\nu ) 
\frac{\alpha_{\overline{MS}}(Q_{1}^{2})}{\pi } \Biggr] ,
\label{kernelact} 
\end{equation}
where 
\begin{equation}
\chi _{L}(\nu )=2\psi (1)-\psi (1/2+i\nu)-\psi (1/2-i\nu)
\label{chil}
\end{equation}
is the function related with the LO eigenvalue, 
$\psi =\Gamma ^{\prime}/\Gamma $ denotes the Euler $\psi $-function, the
$\nu $-variable is conformal weight parameter \cite{Lipatov97}, $N_c$ is
the number of colors, and $Q_{1}$ is the virtuality 
of the reggeized gluon. Moreover, the 
NLO coefficient $r_{\overline{MS}}$ of 
Eq. (\ref{kernelact}) can be decomposed into  a $\beta$-dependent and 
a conformal ($\beta$-independent) part [$
r_{ \overline{MS}} (\nu) = 
r_{ \overline{MS}}^{\beta}(\nu) + r_{ \overline{MS}}^{conf} (\nu)$]
 (For more details see Ref. \cite{NLOBLM}). The  NLO BFKL Pomeron intercept then reads for $N_c = 3$ as follows
\begin{equation}
\alpha_{\pom}^{\overline{MS}} - 1  = 
\omega_{\overline{MS}}(Q^2,0) =
12 \, \ln2 \, \frac{ \alpha_{\overline{MS}}(Q^2)}{\pi} \biggl[ 
1 + r_{\overline{MS}}(0) 
\frac{\alpha_{\overline{MS}}(Q^2)}{\pi} \biggr] \, , 
\end{equation}
with 
\begin{equation}
r_{\overline{MS}}(0) \simeq -20.12 - 0.1020 N_F + 0.06692 \beta_0 ,
\label{rms0}
\end{equation}
and $$r_{\overline{MS}}(0)_{\vert N_F =4} \simeq -19.99 .$$
As discussed in Refs.~\cite{negative} the NLO corrections as given in the $\overline{\mbox{MS}}$-scheme  implies  that the highest eigenvalue of the BFKL equation turns out to be negative and even larger than the LO contribution for $\alpha_s > 0.157$. In Ref. \cite{NLOBLM} the authors have demonstrated that this situation can be improved evaluating the intercept of the BFKL Pomeron at NLO using the BLM scale setting within non-Abelian physical schemes, such as the momentum space subtraction (MOM) scheme. In this case one has
\begin{equation}
\omega_{BLM}^{MOM}(Q^{2},\nu) =  
N_c \chi_{L} (\nu) \frac{\alpha_{MOM}(Q^{MOM \, 2}_{BLM})}{\pi}
\Biggl[1 + 
r_{BLM}^{MOM} (\nu) \frac{\alpha_{MOM}(Q^{MOM \, 2}_{BLM})}{\pi} \Biggr] ,
\end{equation}
where
\begin{equation}
r_{BLM}^{MOM} (\nu) =  r_{ \overline{MS}}^{conf} + T_{MOM}^{conf}
\end{equation}
 and 
\begin{equation}
T_{MOM}^{conf} = \frac{N_c}{8} \biggl[ \frac{17}{2} I + 
\xi \frac{3}{2} (I-1) + \xi^2 (1-\frac{1}{3}I) - 
\xi^3 \frac{1}{6} \biggr]
\end{equation}
with $I=-2 \int^{1}_{0}dx \ln(x)/[x^2-x+1]\simeq 2.3439$ and $\xi$ a gauge parameter.
The $\beta$-dependent part of the $r_{MOM}(\nu)$ defines the
corresponding BLM optimal scale 
\begin{equation}
Q^{MOM \, 2}_{BLM} (\nu) = Q^2 \exp 
\Biggl[ - \frac{4 r_{MOM}^{\beta}(\nu)}{\beta_0} \Biggr] 
= Q^2 \exp \Biggl[ \frac
{1}{2}\chi_L (\nu) - \frac{5}{3} + 2 \biggl(1+\frac{2}{3} I \biggr) \Biggr].
\label{qblm}
\end{equation}
At large $\nu$ one obtains
\begin{equation}
Q^{MOM \, 2}_{BLM} (\nu) = Q^2 \frac{1}{\nu} \exp \biggl[ 2 \biggl(1 + \frac{2}{3}
I \biggr) - \frac{5}{3} \biggr].
\end{equation}
At $\nu=0$ one
has $Q^{MOM \, 2}_{BLM} (0) = Q^2 \bigl( 4 \exp [2(1+2 I /3)-5/3] \bigr) \simeq
Q^2  \, 127$. As a consequence of this procedure, the NLO value for the intercept of the BFKL Pomeron has a very weak dependence on the  virtuality $Q^2$ \cite{NLOBLM}. Moreover, it implies that $(\alpha_{\pom} - 1) = \omega_{NLO} = 0.13 - 0.18$. As emphasized before, using this approach it is possible to describe the  OPAL and L3 data at LEP2 \cite{gammaNLO}.

Another alternative to cure the highly unstable perturbative expansion of the BFKL kernel was proposed in Ref. \cite{ccs}, who realized that the large NLO corrections emerge from the collinearly enhanced physical contributions. A method, the $\omega$-expansion, was then developed to resum collinear effects at all orders in a sistematic way. The resulting RG improved BFKL equation was consistent with  renormalization group requirements through matching to the DGLAP limit and resummation of spurious poles. In this approach the kernel is positive in a much larger region which includes the experimentally accessible one. Based on this approach, but with a fixed coupling,  Khoze, Martin, Ryskin and Stirling (KMRS) \cite{kmrs} proposed recently a simpler model for the NLO corrected LO kernel (See Ref. \cite{kmrs} for details). 
In this approach the position of the singularity (pole) is given by a implicit equation,
\begin{equation}
\omega_{\mathrm{NLO}} = \overline{\alpha}_s \chi(\nu, \omega_{\mathrm{NLO}}) \label{krms}
\end{equation}
where $\overline{\alpha}_s = N_c/\pi\alpha_s$ and the characteristic function $\chi$ is given by
\begin{equation}
\chi(\nu, \omega_{\mathrm{NLO}}) = \chi_0(\nu) + \overline{\alpha}_s \chi_1(\nu, \omega_{\mathrm{NLO}}),
\end{equation}
with $\chi_0(\nu) = \chi_L$ defined in Eq. (\ref{chil}).
The implicit equation, eq. (\ref{krms}), can be rewritten as
\begin{equation}
\omega_{\mathrm{NLO}} = \overline{\alpha}_s\ \left\{ \chi_0^{\mathrm{ht}}(\nu) (1-\omega) + \frac{(1 + \omega A_1(\omega))(1 + \omega)}{(1/2 + \omega/2)^2 + \nu^2} \right\},
\end{equation}
where $\chi_0^{\mathrm{ht}}$ is the higher twist part of $\chi_o$,
\begin{equation}
\chi_0^{\mathrm{ht}}(\nu) = 2\psi(1) - \psi(3/2 - i\nu) - \psi(3/2 + i\nu),
\end{equation}
and $A_1(\omega)$ is obtained from the Mellin transform of the DGLAP splitting function $P_{gg}$, 
\begin{equation}
A_1(\omega) \approx -\frac{11}{12} - \frac{N_F}{18} + \left[\frac{67}{36} - \frac{\pi^2}{6}\right]\omega.
\end{equation}
In our calculations we neglect the effect of the quarks ($N_F = 0$) and solve  the Eq. (\ref{krms}) numerically.  In what follows we present for the first time the predictions for the total $\gamma^* \gamma^*$ cross section considering this improved NLO kernel.

The next-to-leading order corrections discussed above are only part of the corrections to the total $\gamma^* \gamma^*$ cross section. In a full calculation one should also consider the NLO corrections to the impact factors. In the last few years, the real and virtual corrections which contributes at NLO has been estimated \cite{bartelsnlo} and  recently the first numerical results were presented \cite{grigorios}. These preliminary results indicate that the NLO corrections  tend to decrease the value of the impact factors. 
In the phenomenological analyzes what follows we will assume that the main energy-dependent NLO corrections comes from the NLO BFKL kernel rather than from the photon impact factors. Consequently,  our estimates for the total cross section at NLO should be consider an upper bound. We will use in our NLO BFKL calculations the impact factors as given in Eqs. (\ref{eqn:impactT}) and (\ref{eqn:impactL}) and $\omega(Q^2,\nu)$ taken in the NLO. Moreover, following \cite{gammaNLO} we consider the Yennie gauge, where $\xi = 3$.

%\section{Basic Formulas}

%\subsection{$\gamma^* \gamma^*$ total cross section in the dipole picture}
 
Let us now introduce the main formulas concerning the two-photon interactions in the color dipole picture.  At high energies, the scattering process can be seen
 as a succession on time of two
factorizable subprocesses: i) the photon fluctuates in 
quark-antiquark pairs (the dipoles), ii) these color dipoles interact and produce the final state.
The
corresponding cross section is given by
\begin{eqnarray}
\sigma^{\gamma^* \gamma^*}(s_{\gamma\gamma},Q_1^2,Q_2^2) &=&  \sum_{i,j} \sum_{a,b = 1}^{n_f} \int dz_1 \,dz_2 d^2 \,\rkr_1 d^2 \rkr_2 \, |\Psi_i^a (\rkr_1, z_1,Q_1^2)|^2 \, \\ \nonumber 
&\times& \sigma_{a,b}^{\mathrm{dd}}\,(x_{12},\rkr_1, \rkr_2)\, |\Psi_j^b (\rkr_2, z_2,Q_2^2)|^2 ,
\label{dipole_formula}
\end{eqnarray}
where $i,j$ are indices associated to the polarization states of the virtual photons and $\Psi_i$  are the light-cone wavefunctions  of the photon.  The variable $\rr_1$ defines the relative transverse
separation of the pair (dipole) and $z_1$ $(1-z_1)$ is the
longitudinal momentum fractions of the quark (antiquark) of flavours $a,b$. Similar definitions are valid for $\rr_2$ and  $z_2$. The $x_{12}$ variable is defined by
\begin{eqnarray}
x_{12}= \frac{Q_1^2 + Q_2^2 + 4 m_a^2 + 4 m_b^2}{s_{\gamma \gamma} + Q_1^2 + Q_2^2} \,\,.
\label{bjorken}
\end{eqnarray}
 The basic
blocks are the photon wavefunction, $\Psi_i$ , and the dipole-dipole  cross
section, $\sigma_{d\,d}$.

In the dipole formalism, the light-cone
 wavefunctions $\Psi_i(z,\,\rr)$ in the mixed
 representation $(\rr,z)$ are obtained through two dimensional Fourier
 transform of the momentum space light-cone wavefunctions
 $\Psi_i(z,\,\rk)$  which are directly related to the impact factors $\Phi_i$ discussed before (see more details, e.g. in Ref. \cite{predazzi}). The
 normalized  light-cone wavefunctions for longitudinally ($L$) and
 transversely ($T$) polarized photons are given by:
\begin{eqnarray}
 |\Psi_T^f(\rkr, z,Q^2)|^2 & = &  \frac{6\,\alpha_{em}}{4\,\pi^2}\,\sum_f \,e_f^2 \, [z^2 + (1-z)^2]\, \varepsilon_f^2\, K_1^2\,(\varepsilon_f\,r) + m_f^2\,  K_0^2\,(\varepsilon_f\,r)\,, \label{transwf}\\
 |\Psi_L^f(\rkr, z,Q^2)|^2 & = &  \frac{6\,\alpha_{em}}{4\,\pi^2}\,\sum_f \,e_f^2 \, 4\,Q^2\,z^2(1-z)^2 \, K_0^2\,(\varepsilon_f\,r)\,,
\end{eqnarray}
where $e_f$ and $m_f$ stand for the charge and  mass of the quark
having flavor $f$ and $K_{0,1}$ are the McDonald-Bessel functions.
As usual, the quantity $\varepsilon$ is defined as
$\varepsilon^2=z(1-z)\,Q^2+m_f^2$.
 The quark mass
$m_f$ plays a role of a regulator when the regime of $Q^2\rightarrow 0$ is reached.  Namely, it prevents non-zero argument for the
modified Bessel functions $K_{0,1}(\varepsilon r)$ towards $Q^2\rightarrow 0$.

In Ref. \cite{Kwien_Motyka}, Timneanu-Kwiecinski-Motyka (TKM) used the saturation model proposed by Golec-Biernat and Wusthoff to describe $ep$ collisions \cite{GBW} and generalized it two-photon interactions at high energies. The basic idea is that the dipole-dipole cross section $\sigma_{dd} (x_{12},\rr_1,\rr_2)$  has the same functional form as the dipole-proton one and  is expressed in terms of an effective radius $\rr_{\mathrm{eff}}$, which depends on $\rr_1$ and/or $\rr_2$. One has that \cite{Kwien_Motyka},
\begin{eqnarray}
\sigma_{dd}^{\mathrm{TKM}} (x_{12}, \,\rr_{\mathrm{eff}})  =  \hat{\sigma}_0 \, \left[\, 1- \exp
\left(-\frac{\,Q_{\mathrm{sat}}^2(x_{12})\,\rr^2_{\mathrm{eff}}}{4} \right) \, \right]\,, \label{gbwdippho}
\end{eqnarray}
where the  $x_{12}$ variable is given by the Eq. (\ref{bjorken}) and $\hat{\sigma}_0 = \frac{2}{3} \sigma_0$, with $\sigma_0$ the same as in Ref. \cite{GBW}.   The last relation can be justified in terms of the quark counting rule. This model  interpolates between the small and large dipole
configurations, providing color transparency behavior, $\sigma_{dip} \sim \rr^2$, at $\rr \ll 1/Q_{\mathrm{sat}}$  and constant behavior at
large dipole separations $\rr > 1/Q_{\mathrm{sat}}$.  The physical scale which characterizes the transition between  the  dilute and  saturated system is denoted  saturation scale,  $Q_{\mathrm{sat}}\propto x^{-\lambda}$, which is energy dependent. 
In Ref. \cite{Kwien_Motyka} three different scenarios for $\rr_{\mathrm{eff}}$ has been considered, with the dipole-dipole cross section presenting in all cases the color transparency property  ($\sigma_{dd} \rightarrow 0$ for $\rr_1 \rightarrow 0$ or $\rr_2 \rightarrow 0$) and saturation ($\sigma_{dd} \rightarrow \hat{\sigma}_0$) for large size dipoles. We quote also Ref. \cite{marquet} for further analytical studies in the role played by different choices for the effective radius.  In what follows, we use the model I from \cite{Kwien_Motyka}, where  $\rr^2_{\mathrm{eff}} = \rr_1^2 \rr_2^2/(\rr_1^2 + \rr_2^2)$, which is favored by the $\gamma^* \gamma^*$ and $F_2^{\gamma}$ data.  Moreover, in order to extend the dipole model to large $x_{12}$ it is necessary to taken into account threshold correction factors which constrain that the cross section vanish when $x_{12} \rightarrow 1$ as a power of $1-x_{12}$. As in Ref. \cite{Kwien_Motyka}, we multiply the dipole-dipole cross section by the factor $(1-x_{12})^5$. A comment is in order here. One shortcoming of the GBW model is that it does not contain the correct DGLAP limit at large virtualities. Consequently, we may expect that its predictions are only valid at small values of the photon virtualities. Therefore, in what follows we only consider photon virtualities up to 15 GeV$^2$.

\begin{figure}[t]
\includegraphics[scale=0.36]{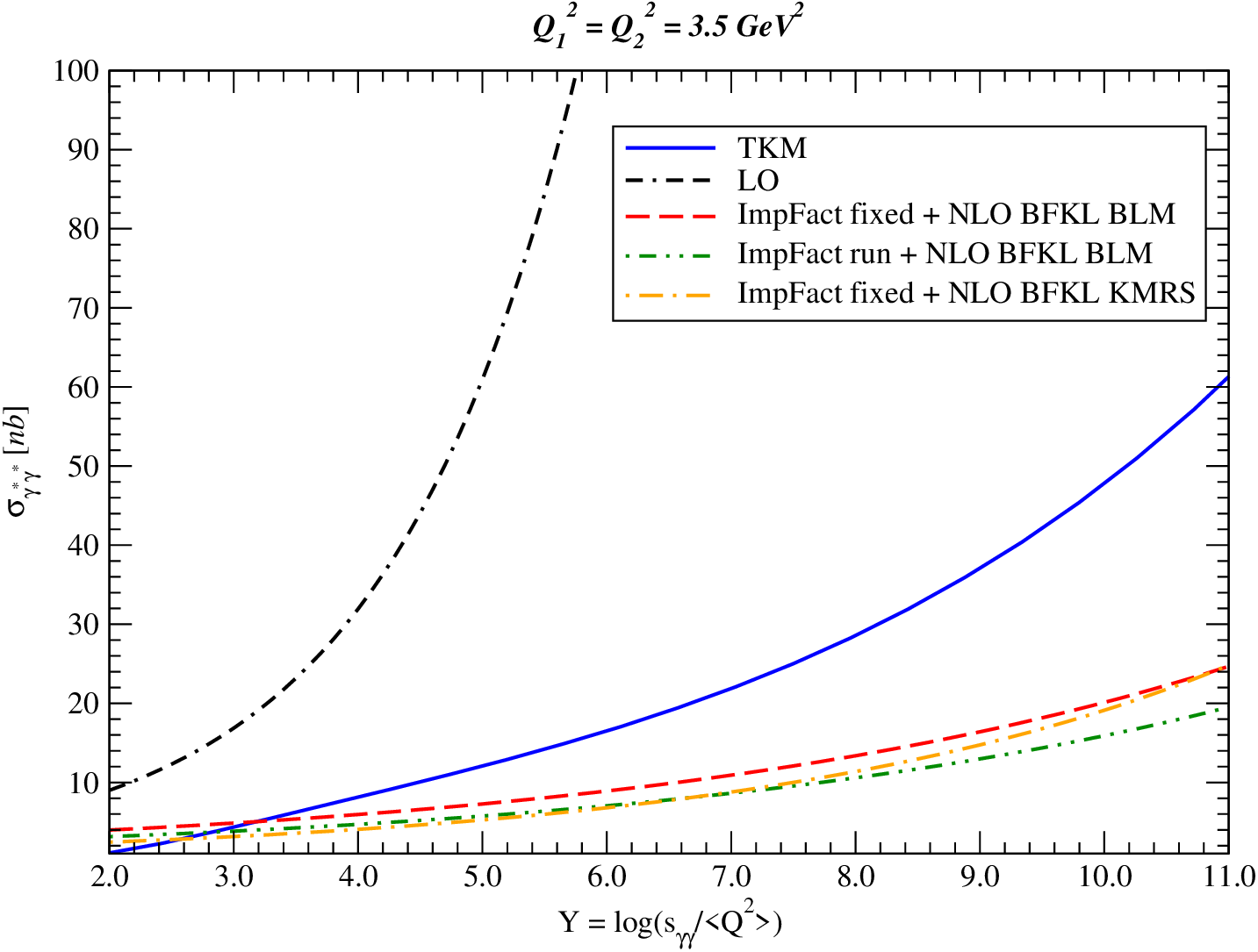}
\caption{Energy dependence of the total $\gamma^* \gamma^*$ cross section  considering distinct  approaches for the QCD dynamics  ($Q_1^2 = Q_2^2 = 3.5$ GeV$^2$).}
\label{fig1}
\end{figure}

In what follows we present our results for the $\gamma^* \gamma^*$ total hadronic cross section. The quark-box contribution is not included in our calculations. Moreover, as the LO impact factors are $\alpha_s$ dependent, we also estimate the effect in our predictions of a running coupling constant. Following Ref. \cite{bartels} we assume in the LO calculation that $\alpha_s = 0.208$. This same value is used in the impact factors in the NLO BFKL calculations. The factor scaling the energy $s$ in the ressumed logarithms, $s_0$, is a free parameter not predicted by the theory. Here we assume that $s_0 = Q^2$, where $Q^2 = \sqrt{Q_1^2 \, Q_2^2}$, and that $Q^2$ is the scale of $\alpha_s$, when the running is considered.

Let us proceed in comparing the different predictions for the total $\gamma^*\gamma^*$ cross section as a function of the variable $Y=\ln(s_{\gamma\gamma}/Q^2)$. For instance, for $Y$ about 10-11, the corresponding $\gamma^*\gamma^*$ center of mass energy reaches 1 TeV, whereas for $Y=7-8$ one has $W_{\gamma \gamma}\sim 500$ GeV. As a first analysis, in Fig \ref{fig1} we consider photons sharing equal virtualities, $Q_1^2 = Q_2^2 = 3.5$ GeV$^2$. The LO BFKL result (double dashed - dot line)  presents a  steep increasing as a consequence of the large effective intercept, $(\alpha_{\pom} - 1) = \frac{4\alpha_s\,N_c}{\pi}\,\ln 2 \approx 0.5$. Clearly, this behavior is ruled out by the L3/OPAL experimental data at high energies \cite{gammaNLO}. For the NLO BFKL approach, the BLM scheme and KMRS approach are considered. We have tested the effect of running coupling constant within the impact factors and verified that the magnitude of the effect is similar in both cases. Consequently, in what follows we only present the predictions considering the running coupling for the BLM scheme.  The fixed $\alpha_s=0.208$ result is labeled by dashed line, whereas the running coupling
 calculation is the double dot-dashed curve. The effect reaches 10\% in the overall normalization at high energies in the case of equal photon virtualities and the energy behavior seems to be unchanged. 
Moreover, we have that the  KMRS prediction (dot-dashed line) is similar to the BLM one at this virtuality. Finally, the solid curve shows the TKM model, which describe the low energy L3/OPAL data ($Y\leq 5$).

\begin{figure}[t]
\begin{tabular}{cc}
\includegraphics[scale=0.36]{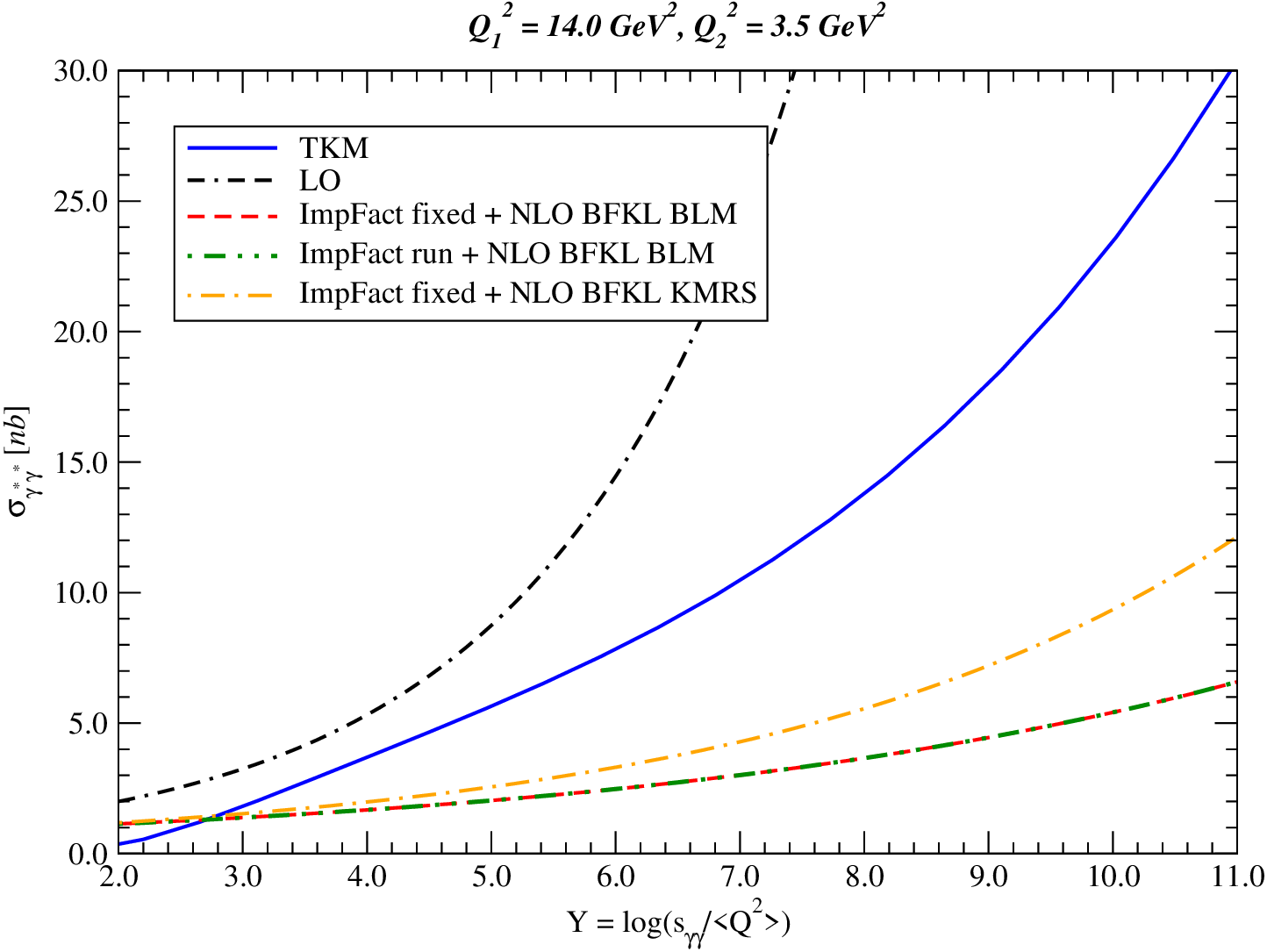} & \includegraphics[scale=0.36]{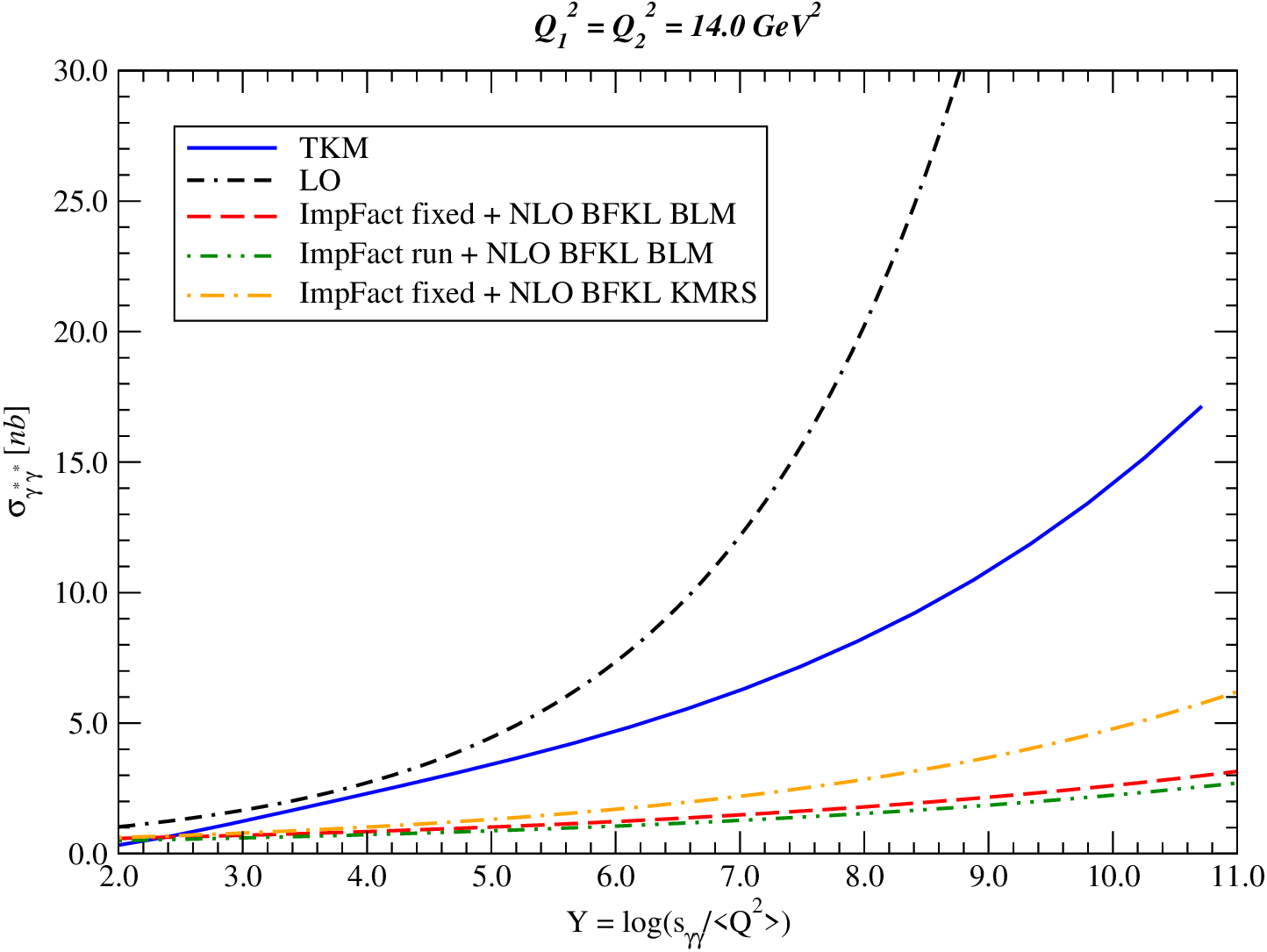} \\
(a) &  (b)
\end{tabular}
\caption{Energy dependence of the total $\gamma^* \gamma^*$ cross section  considering different approaches for the QCD dynamics. (a) $Q_1^2 = 3.5$, $Q_2^2 = 14.0$ GeV$^2$  and (b) $Q_1^2 = Q_2^2 = 14.0$ GeV$^2$.}
\label{fig2}
\end{figure}

The saturation model and the NLO BFKL approach have distinct energy behaviors. The underlying dynamics could be disentangled  whether a sufficiently large energy collider to be available. For instance, at $W_{\gamma\gamma}\simeq 500$ GeV the deviation is a factor 2 and reaches a factor 3 at $W_{\gamma\gamma}\simeq 1$ TeV. The NLO BFKL has a flatter energy growth in contrast with the saturation model. Roughly, at high energies one has $\sigma \propto (s_{\gamma\gamma}/Q_1Q_2)^{0.25}$ for the TKM model and $\sigma \propto (s_{\gamma\gamma}/Q_1Q_2)^{0.2}$ for NLO BFKL with fixed coupling constant in impact factor. An important remark is that although the energy  growth of the total cross section predicted by the BFKL equation is slower when the NLO corrections are included, its solution still present the diffusion property, i.e. the random walk in the transverse momentum. As the mean width of this random walk increases as $\sqrt{\ln s}$, we have that at large $s$ the solution eventually enters into the non-perturbative region. Thus, independently how large are the transverse scales of the scattering objects, there is always an energy beyond which perturbative theory loses its predictive power.

In Fig. \ref{fig2}, different configurations for photon virtualities are taken into account: (a) the assymetric case $Q_1^2 = 3.5$ and $Q_2^2 = 14.0$ GeV$^2$  and (b) the symmetric case $Q_1^2 = Q_2^2 = 14.0$ GeV$^2$ with a somewhat larger virtuality. In both cases the deviations are stronger than the previous analysis. At $Y\simeq 11$ the saturation  model can reach a factor 6 above the NLO BFKL (BLM)  in plot (a) and a factor 7 in plot (b). Notice the small effect of running coupling constant in NLO BFKL when at last one of virtualities is large. The effective power for the saturation model has increased in such way that $\sigma \propto (s_{\gamma\gamma}/Q_1Q_2)^{0.27}$ and for the NLO BFKL (BLM) it remains unchanged. Therefore, the difference between the BFKL NLO (BLM) and TKM predictions grows with $Q^2$. In contrast, the virtuality dependence obtained using the KMRS approach is distinct, which implies a steeper energy dependence for the NLO BFKL (KMRS) prediction. In this case, we have that at $Y\simeq 11$ the saturation  model can reach a factor 2.5 above the NLO BFKL (KMRS)  in plot (a) and a factor 3 in plot (b). Concerning the saturation model, we have also checked that the main contribution comes from the small size dipole configurations where $\sigma_{dd}\propto r_{\mathrm{eff}}^2$ and saturation becomes incleasingly important at $W_{\gamma \gamma}\simeq 1$ TeV. Therefore, these effects must can be strong for the photon structure funtion, $F^{\gamma}_2(x,Q^2)$, where one of photons is real. 

As a conclusion, the results presented here will be important at possible future $e^+e^-$ colliders (TESLA,CLIC), where the available energies might reach 500 GeV or even 1 TeV. In that regime the underlying dynamics could be disentangled. On the other hand, it is interesting the comparison of the present results with those ones using BFKL approach corrected by sub-leading effects \cite{motyka} (where the energy behavior seems to be similar to TKM model) and Regge phenomenology for $\gamma^*\gamma^*$ interactions as the two-Pomeron models \cite{dona_dosch}.
%\section{Results}

%\section{Summary}

\section*{Acknowledgments}
 VPG thank G. Chachamis and A. Sabio Vera for useful discussions. This work was partially financed by the Brazilian funding agencies CNPq and FAPERGS.

\end{document}